\documentclass[10pt,conference]{IEEEtran}
\pdfoutput=1
\usepackage{cite}
\usepackage{amsmath,amssymb,amsfonts}
\usepackage{algorithmic}
\usepackage{graphicx}
\usepackage{textcomp}
\usepackage{xcolor}
\usepackage{xfrac}
\usepackage[font={normal,it}]{caption}

\def\BibTeX{{\rm B\kern-.05em{\sc i\kern-.025em b}\kern-.08em
    T\kern-.1667em\lower.7ex\hbox{E}\kern-.125emX}}

\begin{document}

\title{Probability-Reduction of Geolocation using Reconfigurable Intelligent Surface Reflections}
\author{Anders M. Buvarp$^{1}$, Daniel J. Jakubisin$^{1}$, William C. Headley$^{1}$ and Jeffrey H. Reed$^{2}$ \\ 
Virginia Tech National Security Institute$^1$ and Wireless@VT$^2$ \\
Blacksburg, VA 24061, USA \\
Email: \{abuv,djj,cheadley,reedjh\}@vt.edu. 
}

\maketitle

\begin{abstract}

With the recent introduction of electromagnetic meta-surfaces and reconfigurable intelligent surfaces, a paradigm shift is currently taking place in the world of wireless communications and related industries.  These new technologies are of great interest as we transition from the $5^{th}$ generation mobile network (5G-NR) towards the $6^{th}$ generation mobile system standard (6G).  In this paper, we explore the possibility of using a reconfigurable intelligent surface in order to disrupt the ability of an unintended receiver to geolocate the source of transmitted signals in a 5G-NR communication system.  We investigate how the performance of the \emph{Multiple Signal Classification} (MUSIC) algorithm at the unintended receiver is degraded by correlated reflected signals introduced by a reconfigurable intelligent surface in the wireless channel. We analyze the impact of the direction of arrival, delay, correlation, and strength of the reconfigurable intelligent surface signal with respect to the line-of-sight path from the transmitter to the unintended receiver.  An effective method is introduced for defeating direction-finding efforts using dual sets of surface reflections. This novel method is called \emph{Geolocation-Probability Reduction using dual Reconfigurable Intelligent Surfaces} (GPRIS).  We also show that the efficiency of this method is highly dependent on the geometry, that is, the placement of the reconfigurable intelligent surface relative to the unintended receiver and the transmitter.

\end{abstract}

\begin{IEEEkeywords}
Reconfigurable intelligent surfaces, MUSIC, low probability of geolocation, direction-finding, 5G-NR, smart radio environments.
\end{IEEEkeywords}

\vspace{-3mm}
\section{Introduction}

Until recently, designers of wireless communications systems have considered the radio propagation channel as outside of the control of  communication system design. However, Reconfigurable Intelligent Surfaces (RIS) present the opportunity to introduce control over the propagation channel itself \cite{renzo2020}.  Thus, RIS is a paradigm shift from previous generation systems that has the potential to play a significant role in the next generation of wireless systems such as 6G. A related and developing field of great interest is Electromagnetic Signal and Information Theory (ESIT) and Holographic Radios \cite{dardari2021}.

Ongoing RIS research includes redirecting signals away from adversarial nodes or redirecting a signal in a desired direction, and hence reducing the overall interference \cite{renzo2020}.  In this work, we explore the use of a RIS to disrupt the direction-finding (DF) capability of a potentially adversarial receiver that might want to jam a friendly communication link.  Disruption of DF is a key capability for \emph{Low Probability of Geolocation} (LPG) in millimeter wave systems.  LPG is important in military applications where detection, isolation, and identification of signals must be avoided.

\begin{figure}[htbp!]
    \centering
    \includegraphics[width=0.8\columnwidth]{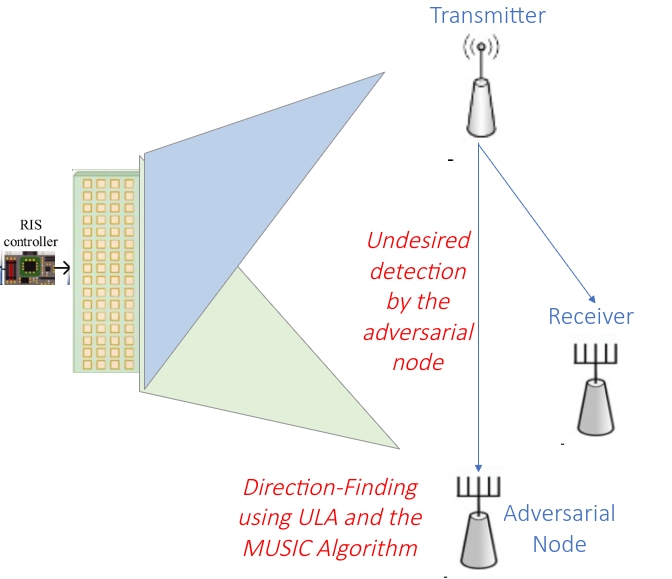}
    \vspace{-2mm}
    \caption{A signal transmitted while an adversarial node uses the MUSIC algorithm for geolocation of the transmitter for signal detection and jamming.  A RIS is employed to reflect a range of correlated multipath signals to defeat the detection.}
    \label{fig:orientation}
\end{figure}

In certain scenarios, a transmitter will not be able to fully avoid radiating signal towards an adversarial node, e.g., due to the node geometry or transmit array limitations. We propose using a RIS in order to introduce artificial correlated multipath into the environment to prevent geolocation.  Fig. \ref{fig:orientation} shows an adversarial node attempting to geolocate a transmitter using a direction-finding algorithm called \emph{Multiple Signal Classification} (MUSIC) and a RIS that is reflecting obfuscating signals towards the Uniform Linear Array (ULA) of the adversary.  The contribution of this paper is to understand the potential for this technique to impact DF under certain geometrical constraints and to motivate future work on practical implementations. Along these lines, we assume knowledge of the adversarial node's location and accurate control of the RIS element phases. 
\vspace{-5mm}

There are several Direction-of-Arrival (DOA) estimation methods that could be used for geolocation e.g. Doppler DF \cite{bai2019}, Watson-Watt \cite{oestreich2012}, Correlative Interferometry \cite{gou2018}, Time Difference of Arrival (TDOA) \cite{zang2020}, Maximum Likelihood Estimation (MLE) \cite{emery2016}, MUSIC \cite{gu2003}, Estimation of Signal Parameters via Rotational Invariance Techniques (ESPRIT) \cite{gong2008}, and Matrix Pencil \cite{aytas2021}.  MLE, MUSIC and ESPRIT all depend on estimating the correlation matrix, $R$, using $K$ snapshot samples of the incoming signal received by an antenna array with $N$ elements.  The output of these algorithms is the signal directions, $\Theta_i$, of the $M$ incoming signals. The Matrix Pencil algorithm is similar to ESPRIT, however it is a non-statistical method.

The advantages of the MUSIC algorithm compared to classical DOA estimation methods include general sensor configurations, ultra-slow sampling, and small array dimensions. MUSIC can be applied to both narrowband and wideband signals without prior knowledge of the signals \cite{gustaf2015}.  The MUSIC algorithm, which is closely related to Prony's method \cite{prony1795}, has been evaluated as having superior high-resolution performance at the cost of computation and storage \cite{barabell1998}. Using peaks of a \emph{spatial} pseudo-spectrum, MUSIC is able to identify the angle-of-arrival of signals from $-90$ to $+90$ degrees. For this reason, we chose to evaluate our DF disruption technique against an adversarial node equipped with the MUSIC algorithm.

This paper is organized as follows. Section \ref{ris_model} defines the RIS wave propagation model. How to program the surface to maximize the Signal-to-Noise Ratio (SNR) at the adversarial node is explained in Section \ref{array_proc}. In Section \ref{results}, we evaluate the impact of correlated multipath on MUSIC direction-finding performance, independent on the source of the multipath. In Section \ref{obfuscation} the RIS propagation model is applied and performance is evaluated as a function of RIS and node geometry. We end the paper with conclusions in Section \ref{conclusions}.

\section{RIS Wave Propagation Model}
\label{ris_model}

A RIS is a type of  electromagnetic metasurface \cite{renzo2020}.  Typically, it is a sheet of inexpensive and adaptive thin composite material, which can cover walls, buildings, ceilings, etc.  It consists of individually tuned passive reflectors, making the phase response of the incoming wave tunable.  The individually controlled unit cell of a RIS incorporates low power electronic circuits with components such as positive-intrinsic-negative (PIN) diodes and varactors, where the bias voltage of the varactor can be tuned.  The RIS can be controlled and programmed using a simple microcontroller such as Raspberry Pi. The RIS is reconfigurable and can be programmed to control and modify the incident radio waves by elementary electromagnetic functions such as reflection, refraction, absorption, focusing/beamforming, polarization, splitting, analog processing and collimation \cite{renzo2020}.  A RIS can also be used for joint transmit and passive beamforming design \cite{liu2021}. 

For the direct Line-of-Sight (LOS) path between the RIS and the adversarial node, the attenuation and the delay of the received baseband signal are \cite{matt2021},
\begin{equation} \label{equ:los_amp_dly}
    A_0 = \frac{\lambda _c \sqrt{G_{Tx}^{Adv}  \ G_{Adv}^{Tx}}}{4\pi \lVert \mathbf{p}_{Adv}-\mathbf{p}_{Tx} \rVert} , \quad \tau_0 = \frac{\lVert \mathbf{p}_{Adv}-\mathbf{p}_{Tx} \rVert}{c_0},
\end{equation}
\noindent respectively, where $\lambda _c$ is the wavelength of the carrier wave,  $G_{Tx}^{Adv}$ is the antenna gain of the transmit antenna in the direction of the adversarial node, $G_{Adv}^{Tx}$ is the antenna gain of the adversarial node in the direction of the transmitter, $\mathbf{p}_{Adv}$ is the 3-dimensional (3D) position of the adversarial node, $\mathbf{p}_{Tx}$ is the 3D position of the transmitter, $c_0$ is the speed of light through air, and $\lVert \cdot \rVert$ the $l^2$-norm.  For the RIS path, the attenuation and delay are,
\begin{equation} \label{equ:ris_amp}
    A_{i,j} = \sqrt{\mu}\frac{\lambda _c^2}{16\pi^2}\frac{\sqrt{G_{i,j}^{Adv} \ G_{Adv}^{i,j} \ G_{Tx}^{i,j} \ G_{i,j}^{Tx} }}
    {\lVert \mathbf{p}_{Adv}-\mathbf{p}_{i,j} \rVert
    \lVert \mathbf{p}_{i,j}-\mathbf{p}_{Tx} \rVert} ,
\end{equation}
\begin{equation} \label{equ:ris_dly}
     \quad \tau_{i,j} = \frac{\lVert \mathbf{p}_{i,j}-\mathbf{p}_{Tx} \rVert + \lVert \mathbf{p}_{Adv}-\mathbf{p}_{i,j} \rVert}{c_0},
\end{equation}
\noindent respectively, where ${i,j}$ are the element index, $G_{i,j}^{Adv}$ is the antenna gain of the RIS element in the direction of the adversarial node, $G_{Adv}^{i,j}$ is the antenna gain of the adversarial node in the direction of the  RIS element, $G_{Tx}^{i,j}$ is the antenna gain of the transmitter in the direction of the RIS element,  $G_{i,j}^{Tx}$ is the antenna gain of the RIS element in the direction of the transmitter, $\mu \in \mathcal [0,1]$ is the fraction of the incident energy that is scattered, and $\mathbf{p}_{i,j}$ is the RIS element location with $x,y,z$-coordinates,
\begin{align} \label{equ:pmn_coordinates}
\begin{split}
    x & = 0 \\
    y & = id_y - 0.5d_y((Q+1) \bmod 2) \quad \forall i = [1,Q] \\
    z & = jd_z - 0.5d_z((P+1) \bmod 2) \quad \forall j = [1,P],
\end{split}
\end{align}
\noindent where $d_y$ is the element spacing in the y-direction, $d_z$ is the element spacing in the z-direction, Q is the number of element columns of the surface and P is the number of rows of elements.  The complex baseband signal received at the adversarial node, $y(t)$, can be expressed as,
\begin{equation} \label{equ:rx_baseband}
\begin{split}
    y(t) & = A_0 e^{-j2\pi f_c\tau_0} x(t-\tau_0 ) \quad \\
    & + \sum_{i,j} A_{i,j}   e^{-j2\pi f_c\tau_{i,j} -j\phi _{i,j} }
    x \left( t - \tau_{i,j}  - \frac{\phi _{i,j} }{2\pi f_c}\right) \\
    & + n(t)
\end{split}
\end{equation}
\noindent where $\tau_0$ is the time delay from the transmitter to the adversarial node, $\tau_{i,j}$ is the time delay from the transmitter to the adversarial node via the RIS elements, $\frac{\phi _{i,j} }{2\pi f_c}$ is a tunable delay that enables the anomalous scattering of the incident radio waves and $n(t)$ is additive white Gaussian noise (AWGN).

\section{RIS Array Processing}
\label{array_proc}

The phases $\phi _{i,j}$ are programmed by setting the bias voltage of the varactor of each surface element.  Maximizing the received power of the baseband signal \eqref{equ:rx_baseband} requires that all terms have the same phase.  In order to maximize the SNR at the ULA, $\phi _{i,j}$ is programmed as,
\begin{equation} \label{equ:phi}
    \phi _{i,j} = 2\pi f_c(\tau_0 - \tau_{i,j}) \bmod 2\pi 
\end{equation}
We assume that the location of the adversarial node is known. 

\section{Simulation Results}
\label{results}

Without using the complex RIS model shown in Section \ref{ris_model}, we first analyzed the behavior of the MUSIC algorithm using orthogonal frequency division multiplexing (OFDM) signals generated as they arrive on the adversarial node's ULA. We looked at the ability of the adversarial node to detect signals in the MUSIC pseduo-spectrum when two correlated or two uncorrelated signals arrive at the ULA from different directions.  We also studied the performance of signal delays.  The distributions of MUSIC pseudo-spectrum peak magnitudes were analyzed in order to produce Receiver Operating Characteristic curves. Distributions of the detected angle-of-arrival for correlated and uncorrelated signals are also shown.

\subsection{Simulated OFDM Signals}

As input to MUSIC, we used OFDM signals with a sample rate of $7.68$ Msamples/s and a sub-carrier spacing of $15$ kHz, which is typical of the $4^{th}$ generation mobile system standard \emph{Long Term Evolution} (4G LTE) and $5^{th}$ generation mobile network (5G-NR) transmissions.  These signals were transmitted through an AWGN channel and received by a Uniform Linear Array with $16$ elements. Each OFDM symbol had $300$ sub-carriers to carry a quadrature phase shift keying (QPSK) signal without pilots.  The payload of $600$ data bits were drawn from a Bernoulli distribution with an equal probability of $0$ and $1$.   The length of each OFDM symbol was $512 + 36 = 548$ I/Q samples, where $36$ is the length of the cyclic prefix and $512$ is the inverse FFT size.


\subsection{Receiver Operating Characteristics}

\begin{figure}[htbp!]
    \vspace{-3mm}
    \centering
    \includegraphics[width=\columnwidth]{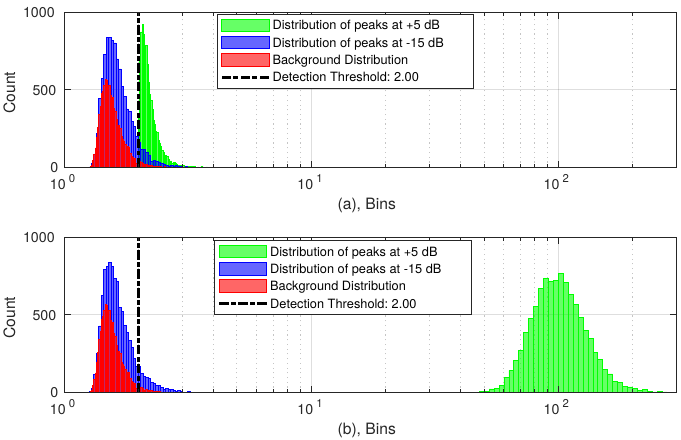}
    \vspace{-5mm}
    \caption{Magnitude distributions of the strongest detected peak for (a) two correlated signals and (b) two uncorrelated signals with arrival angles $\pm30$ dB.}
    \label{fig:histogram_of_peaks}
\end{figure}

Using $10000$ Monte-Carlo iterations, Fig. \ref{fig:histogram_of_peaks} shows distributions of the magnitudes of the pseudo-spectrum peaks for SNR levels $-15$ and $+5$ dB using the OFDM signals, a $16$-element ULA with the LOS azimuth at $30$ degrees with respect to the bore-sight.  We observed that the distributions have a long right-hand tail and a much shorter left-hand tail characteristic of the Rayleigh-distribution.  Note that the background distribution has the same shape as the signal distributions but with a smaller mean.  The background distribution and a detection threshold of $2.0$, corresponding to a $P_{fa}$ of $0.026$, is also shown in Fig. \ref{fig:histogram_of_peaks}. For correlated  $+5$ dB signals, the peaks are much shorter compared to the case of uncorrelated signals.

\vspace{-1mm}

\begin{figure}[htbp!]
    \vspace{-4mm}
    \centering
    \includegraphics[width=\columnwidth]{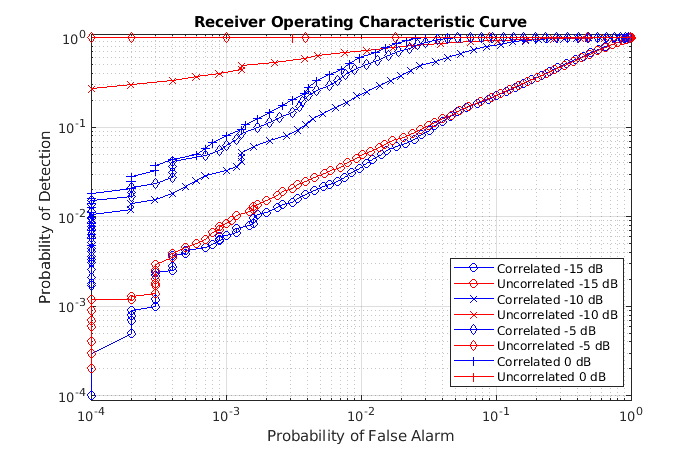}
    \caption{The Receiver Operating Characteristic curves for correlated and uncorrelated OFDM signals.  A detection threshold of $2.0$ corresponds to a $P_{fa}$ of $0.026$.}
    \label{fig:roc}
    \vspace{-4mm}
\end{figure}

Fig. \ref{fig:roc} shows the Receiver Operating Characteristic (ROC), that is, the Probability of Detection, $P_{d}$, as a function of the Probability of False Alarm, $P_{fa}$, for SNR levels in the range from $-15$ to $+5$ dB.  The ROC curves have a knee around $P_{fa} = 0.02$.  After a closer examination of the ROC curves, a detection threshold of $2.0$ was picked corresponding to a $P_{fa}$ of $0.026$ in order to maximize the Probability of Detection while minimizing the Probability of False Alarm.

\subsection{Analysis of Direction-Finding Error vs Azimuth}

\begin{figure}[b!]
    \vspace{-4mm}
    \centering
    \includegraphics[width=\columnwidth]{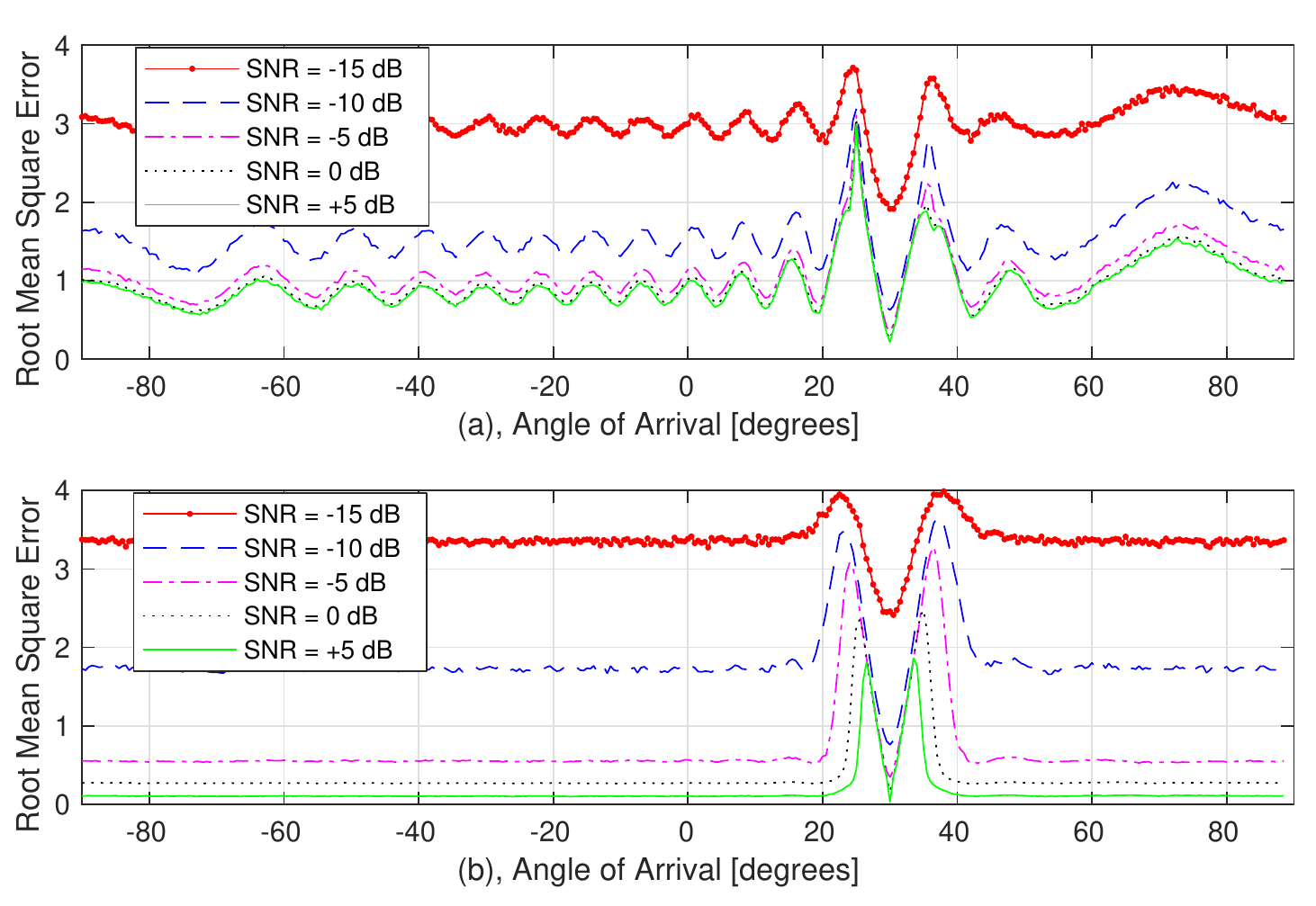}
    \caption{Direction-finding RMS error as a function of the azimuth angle of an additional (a) correlated and (b) uncorrelated OFDM signal. The LOS azimuth angle is $30$ degrees.}
    \label{fig:rmse_vs_aoa}
\end{figure}

From the perspective of the adversarial node, the transmitter was placed at a $30$-degree azimuth angle.  Using a $16$-element ULA and $10000$ Monte Carlo iterations, the DF performance was observed with the azimuth angle of an additional OFDM signal varied from $-90.0$ to $+89.5$ with Signal-to-Noise Ratios (SNR) at $-15, -10, -5, 0, +5$ dB. In order to just observe the impact of the angle-of-arrival, we kept the signals artificially phase coherent, without time delay and with identical receive power. Fig. \ref{fig:rmse_vs_aoa} shows the DF Root-Mean-Square (RMS) error as a function of the azimuth angle of the additional signal.  In general, it can be seen that the RMS error in direction-finding is much lower for uncorrelated signals compared to correlated signals. 
The RMS error is the smallest when the two signals are aligned to the same angle-of-arrival, which makes sense as it would be a simple superposition of two signals. We also conclude that the RMS error is the largest if the difference in angle-of-arrival between the two signals are somewhere around $10$ degrees, even for signals with very low SNR.

\subsection{Analysis of Direction-Finding Error vs Delay}

We know that the signals arriving at the ULA via the RIS path have some additional propagation delay.  Therefore, we decided to investigate the impact of a delay on the direction-finding error. For this purpose, we placed the transmitter at a $30$-degree azimuth angle and an additional OFDM signal with identical receive power at $-30$ degrees.  Using a $16$-element ULA and $10000$ Monte Carlo iterations, the DF performance was observed when the additional signal was delayed from $0$ to $500$ nanoseconds with Signal-to-Noise Ratios (SNR) at $-15, -10, -5, 0, +5$ dB. For correlated signals with SNR $\geq 0$ dB, the RMS error is slightly elevated for very small delays.  We conclude that the correlation coefficient between the arriving signals will be constrained by the geometry of the scenario.

\subsection{Analysis of Number of Peaks Utilized}

\begin{figure}[htbp!]
    \centering
    \includegraphics[width=\columnwidth]{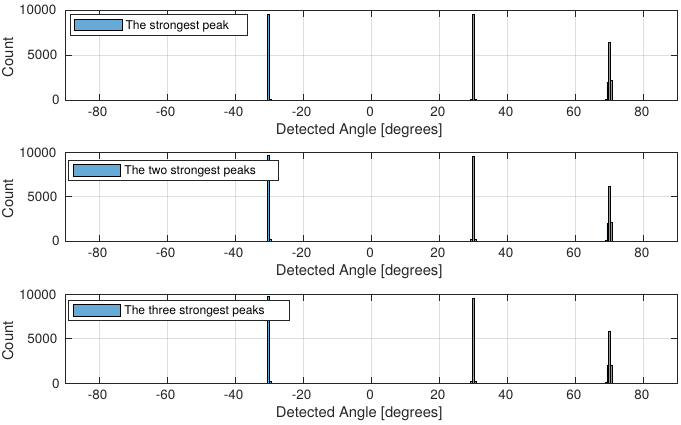}
    \caption{Histogram of $30000$ angles associated with the one, two and three strongest peaks using \emph{uncorrelated} signals. }
    \label{fig:histogram_detected_angles_uncorr}
\end{figure}

The MUSIC algorithm is able to detect the angle-of-arrival of multiple signals by selecting the strongest peaks in the pseudo-spectrum.  An adversarial node might decide to jam all detected signals, hence it is of interest to explore the impact of not only picking the strongest peak but also picking the two, three or any peaks that are observed.  Using a $16$-element ULA, a SNR level of $5$ dB and a detection threshold of $2.0$, we experimented with three signals having the same magnitude and phase and $-30$, $+30$ and $+70$ degrees angle-of-arrival. The MUSIC algorithm is generally successful at direction-finding for uncorrelated signals when selecting the one, two, or three strongest peaks, as seen in Fig. \ref{fig:histogram_detected_angles_uncorr}. 

\begin{figure}[htbp!]
    \centering
    \includegraphics[width=\columnwidth]{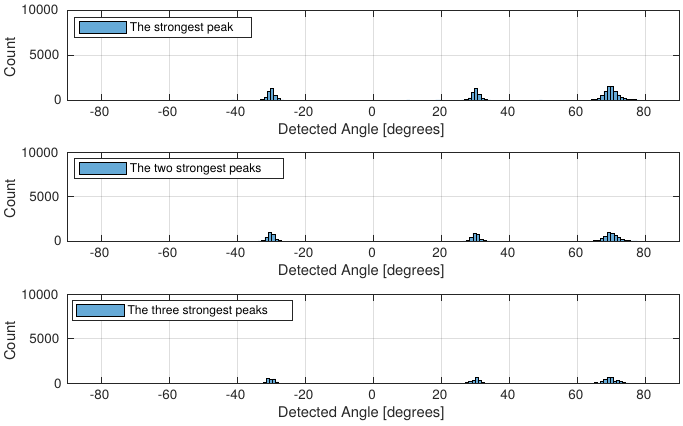}
    \caption{Histogram of $30000$ angles associated with the one, two and three strongest peaks using \emph{correlated} signals. Rate of peaks not found above the threshold: 50\%, 68\% and 78\% respectively. }
    \label{fig:histogram_detected_angles_corr}
    \vspace{-4mm}
\end{figure}

Using the same arrival angles, Fig. \ref{fig:histogram_detected_angles_corr} shows that when the waveforms are correlated, the performance of the MUSIC algorithm is significantly degraded with very small Gaussian-like distributions centered at the expected angles and extremely long tails scattered over the entire pseudo-spectrum.  MUSIC is rarely able to detect the correct angle-of-arrival when the signals are correlated, as observed in the lower total count.

\subsection{Probability of Detection}

\begin{figure}[htbp!]
    \vspace{-4mm}
    \includegraphics[width=\columnwidth]{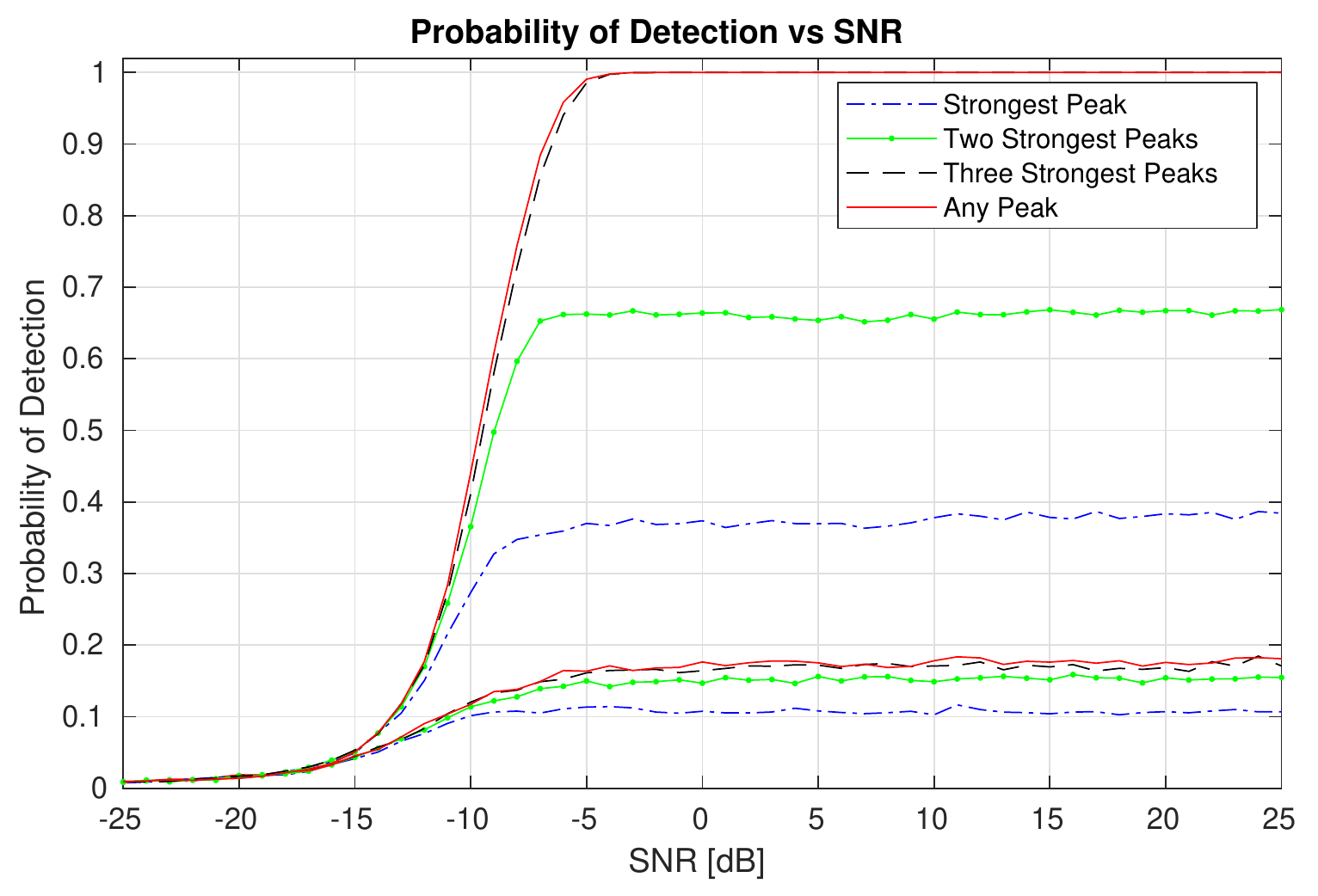}
    \vspace{-4mm}
    \caption{Probability of detection for a ULA with $16$ elements. The lower four curves are for correlated signals and top four are for uncorrelated signals.}
    \vspace{-6mm}
    \label{fig:probability}
\end{figure}

Fig. \ref{fig:probability} shows the probability of detection when selecting the one, two, three strongest peaks or any peak above the detection threshold of $2.0$ within the range of $\pm2.5$ degrees of the LOS azimuth and a Probability of False alarm, $P_{fa} = 0.026$. This was carried out for three correlated and three uncorrelated OFDM signals using $10000$ Monte Carlo iterations. Considering a geometry of a equilateral triangle with the boresight of the ULA splitting the distance between the transmitter and the RIS, the LOS azimuth was set to $+30$ degrees and the additional OFDM signals was set at $-30$.  A third auxiliary signal was added at $+70$ degrees representing incidental multipath in the environment.  For uncorrelated signals above $-5$ dB SNR, the probability of detection goes towards $\sfrac{1}{3}$, if the strongest peak is picked. If two peaks are picked, the probability of detection approaches $\sfrac{2}{3}$ and the probability of detection is close to $1.0$, if three peaks are used. If any peak is above the detection threshold and within $\pm2.5$ degrees of the LOS azimuth, the probability of detection is slightly higher than the case of the three strongest peaks in the uncorrelated case. The MUSIC algorithm is based on that the spatial correlation matrix has a rank that equals the number of signals present.  The probability of detection is poor for correlated signals because the correlation matrix is rank-deficient, which results in weak or non-existent peaks.  Above $-10$ dB SNR, the probability of detection is about $0.1$, if the strongest peak is selected. If additional peaks are selected, the probability of detection is slightly higher but still poor.  

\section{Direction-Finding Obfuscation using a RIS}
\label{obfuscation}

The MUSIC algorithm is based on the assumption that the incident signals are uncorrelated. Here we introduce the complex RIS model shown in Section \ref{ris_model} into our MATLAB simulations in order to explore the performance of RIS-generated multipath.  As we have shown in the previous section, the MUSIC algorithm breaks down in the presence of correlated signals.  Using a RIS, we are able to generate reflections towards the adversarial ULA. These reflections are correlated but delayed versions of the signal from the transmitter.   In order to obfuscate the direction-finding algorithm, we developed a novel method called \emph{Geolocation-Probability Reduction using Dual Reconfigurable Intelligent Surfaces} (GPRIS).  This method produces two correlated and delayed signals via reflections in a \textbf{rectangular-shaped RIS}, resulting in a total of three correlated signals arriving at the adversarial ULA.  The two sets of reflections from each end of the RIS were achieved by programming the phase of the center elements of the RIS in a checkered fashion; the neighbors to the immediate sides, above and below are out of phase by $180$ degrees.  The cancellation of the contributions from the reflections created an effective subdivision of the RIS into two parts.  \emph{An equivalent alternative would be the use of two RIS surfaces}.  The remaining elements at each RIS end were programmed according to Eqn. (\ref{equ:phi}).  By increasing the number of elements at each end, the impact of the reflections on the ULA increased.

\subsection{RIS Model Simulations}
For our simulations using the RIS model, we used a RIS with $5000$ columns and $500$ rows with an element spacing of $\lambda_c/5$ in both directions. The number of rows and columns were picked via experimentation. They were selected large enough to generate reflections that produced signals with angles-of-arrival at the ULA that defied MUSIC.  The RIS was placed at the origin, centered on the X-axis and raised in the direction of the Z-axis. We assumed a millimeter wave system With a $f_c$ of $26$ GHz. The ULA at the adversarial node had $16$ elements. The gain in all directions were set to $0$ dBi. 

\begin{figure}[htbp!]
    \vspace{-4mm}
    \centering
    \includegraphics[width=\columnwidth]{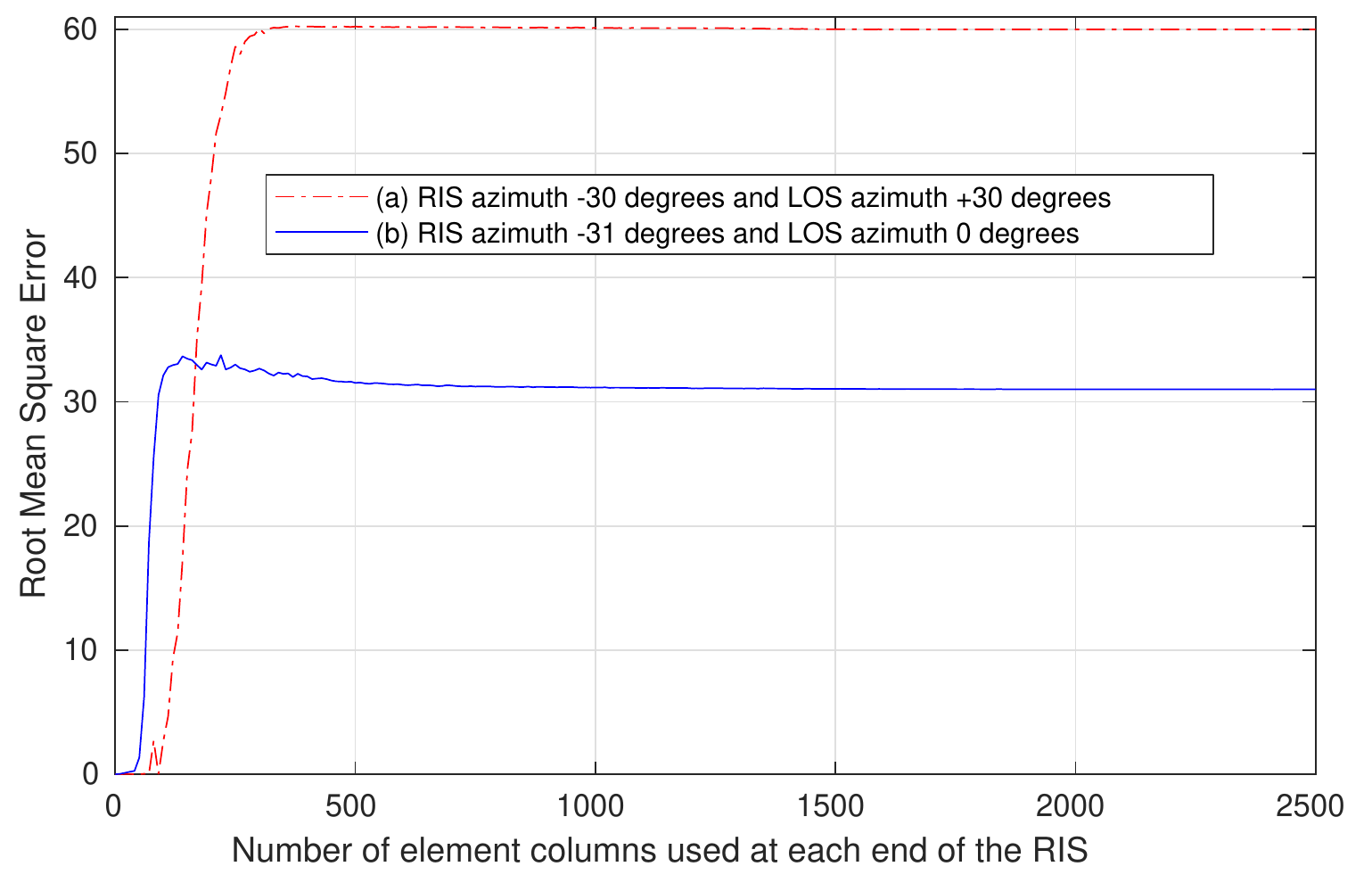}
    \vspace{-5mm}
    \caption{DF RMS error as a function of RIS utilization at each end of the surface; DF performed with a 16-element ULA and $5$ dB OFDM signals. (a) $100$-meter equilateral triangle and (b) triangular geometry with the LOS at the ULA boresight.}
    \label{fig:df_error_ris_model}
\end{figure}

The transmitter was arbitrarily placed at XY-coordinates $(100,0)$ meters and the ULA of the adversarial node was placed at $(50,-86.6)$ meters forming an equilateral triangle together with the center of the RIS and with the vertices $100$ meters apart, that is, the boresight of the ULA was at an equal distance between the RIS and the transmitter. Fig. \ref{fig:df_error_ris_model} shows the resulting RMS Error vs Element Utilization at $5$ dB SNR using only $500$ Monte Carlo iterations, which was due to the processing time of the enormous amount of reflections.  For the first $100$ element columns, MUSIC is only picking up the LOS signal. After that, the error increases as more surface elements are utilized until about $300$ columns of elements Using a detection threshold of $2.0$ with the LOS azimuth at $+30$ degrees and with two additional OFDM signals at $-30$ and $+70$ degrees, at which point MUSIC is only detecting the reflections from the RIS, and the LOS is obfuscated.

Next, another geometry was explored in order to observe the impact on direction-finding obfuscation. The transmitter was placed at XY-coordinates $(30,50)$ and the ULA of the adversarial node was placed at $(30,-50)$ making the LOS azimuth the same as the boresight of the ULA. Fig. \ref{fig:df_error_ris_model} shows that for the first $100$ element columns, MUSIC is only picking up the LOS signal. After about $100$ element columns, MUSIC is only detecting the reflections from the RIS. The former geometry involves longer delays resulting in a smaller correlation coefficient and hence a larger number of element columns are needed to obfuscate the MUSIC algorithm.

\begin{table*}[t]
\caption{\label{tab:rho} Correlation coefficients, $\rho$, for two RIS geometries.}
\centering
\begin{tabular}{l|c|c|c|c|c|c}
Case & Description & Transmitter (x,y) & ULA (x,y) & LOS Power & RIS Power & $\rho$ \\\hline
(a) & Equilateral & $(100,0)$ & $(50,-86.6)$  & $2.14 \times 10^{-6}$ & $1.38 \times 10^{-5}$ & $0.625$\\
(b) & LOS at Boresight & $(30,50)$ & $(30,-50)$  & $2.95 \times 10^{-6}$ & $1.80 \times 10^{-4}$ & $0.889$ \\
\end{tabular}
\end{table*}

\begin{figure}[htbp!]
    \centering
    \includegraphics[width=\columnwidth]{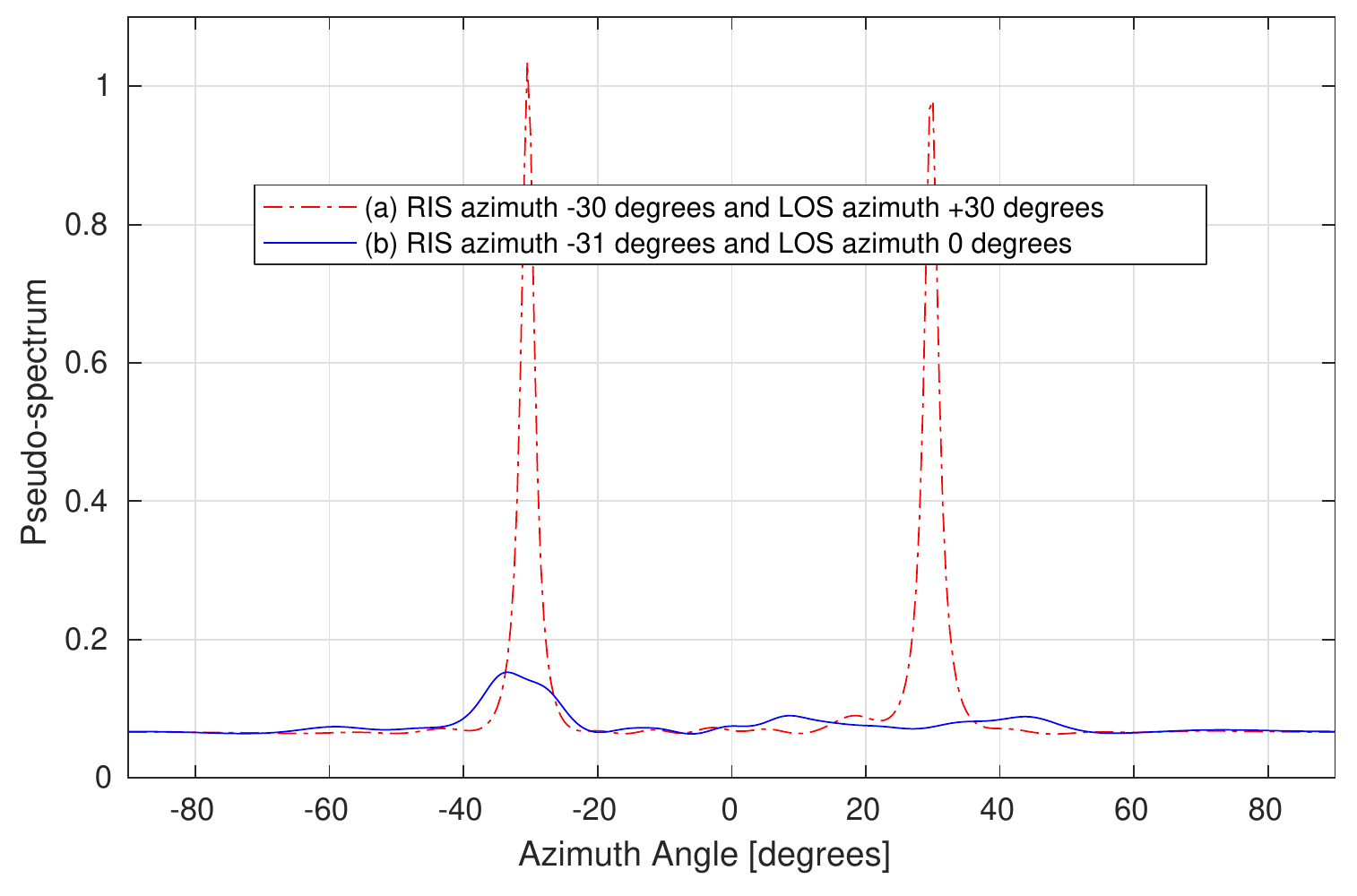}
    \caption{Pseudo-spectrum with three correlated $5$ dB signals for two geometries; (a) equilateral triangle; and (b) triangular geometry with the LOS at the ULA boresight.}
    \vspace{-2mm}
    \label{fig:pseudo-spectrum}
\end{figure}

Using  $300$ columns of elements at each RIS end, Fig. \ref{fig:pseudo-spectrum} shows the pseudo-spectrum of the two different geometries and the fidelity of the associated peaks. Table \ref{tab:rho} shows the correlation coefficients associated with these geometries.  The received signal strengths for the LOS and RIS paths are also included.  The difference in the power of the LOS path in the two aforementioned geometry cases is due to the reduced propagation path.  The larger correlation coefficient in the latter case is due to the significantly reduced relative propagation path lengths.

When the signals of the RIS path and the LOS path are loosely correlated, the MUSIC algorithm is capable of discerning the  angles-of-arrival from two distinct peaks in the pseudo-spectrum, although the fidelity of those peaks are nowhere near that of peaks associated with uncorrelated signals. Vice versa, when the signals of the RIS path and the LOS path are strongly correlated, the pseudo-spectrum devolves into a state of very low fidelity, making it impossible to resolve the angles of the signals impinging on the ULA. Therefore, it is apparent that the geometry greatly influences the ability to defeat the MUSIC algorithm using a RIS.
\section{Conclusions}
\label{conclusions}

The main contribution of this paper is to understand the potential for using a RIS to obfuscate adversarial direction-finding efforts under certain geometrical constraints, and to motivate future work on practical implementations.

In order to minimize the likelihood of accurate direction-finding and hence geolocation and signal detection, it is best to generate correlated signals at the adversarial node. In Fig. \ref{fig:roc} and  Fig. \ref{fig:probability}, we observe an order of magnitude reduction in detection probability for correlated signals, compared with detection of similarly situated uncorrelated signals. 

By programming the phase of the reflections from the RIS elements, we were able to produce two sets of RIS reflections that are correlated with the LOS signal and hence effectively defeat the MUSIC algorithm at the adversarial node, resulting in a significant reduction of the probability of geolocation and adversarial signal detection.  In our simulations, we found that roughly $500$-by-$300$  elements at each end of the RIS were needed to render the MUSIC algorithm completely blind using the GPRIS method, for a millimeter wave system.

The level of correlation between the LOS and RIS paths is largely dependent on the geometry of the scenario.  Geometries which cause a larger propagation delay to the RIS path relative to the LOS path result in reduced correlation between the paths.  In our current work, this puts a constraint on the range of possible locations for the RIS to be effective in disrupting direction finding. 

\section{Acknowledgments}
This work was carried out under U.S. Army SBIR Phase II Contract No. W911NF-21-C-0015.  We would like to thank Intelligent Automation Inc., Rockville, MD for fully supporting this effort.

\bibliographystyle{IEEEtran}
\bibliography{ris.bib}

\end{document}